\documentclass[conference]{IEEEtran}
\IEEEoverridecommandlockouts

\usepackage{cite}
\usepackage{amsmath,amssymb,mathtools}
\usepackage{graphicx}
\usepackage{tabularray}
\UseTblrLibrary{booktabs}
\usepackage{siunitx}
\usepackage[font=small]{caption}
\usepackage{bm}
\usepackage{verbatim}
\usepackage{graphicx}
\graphicspath{{images/}} 
\usepackage{epstopdf}
\usepackage{subfig}
\usepackage{float}
\usepackage{tikz}
\usepackage{tcolorbox}
\usetikzlibrary{arrows.meta, positioning}

\newcommand{\pderiv}[2]{\frac{\partial #1}{\partial #2}}

\newcommand{\deriv}[2]{\frac{d #1}{d #2}}

\makeatletter
\newcommand{\newlineauthors}{%
  \end{@IEEEauthorhalign}\hfill\mbox{}\par
  \mbox{}\hfill\begin{@IEEEauthorhalign}
}
\makeatother

\title{A 2D Axisymmetric Multi-Domain DC Arc Model for Simulink Implementation}
\author{Your Name}
\author{\IEEEauthorblockN{1\textsuperscript{st} Vinod Kumar Maddineni}
\IEEEauthorblockA{\textit{Sr. Power Systems Engineer} \\
\textit{Schneider Electric}\\
California, USA \\
0009-0001-2742-6429}
\and

\IEEEauthorblockN{\hspace{0.5cm}2\textsuperscript{nd} Rahul Reddy Devarapally}
\IEEEauthorblockA{\hspace{0.5cm}\textit{Sr. Electrical Engineer} \\
\textit{K\&A Engineering}\\
North Carolina, USA \\
0009-0005-2597-771X}
\and
\IEEEauthorblockN{3\textsuperscript{rd} Nihar Panchal}
\IEEEauthorblockA{\textit{Sr. Electrical Engineer} \\
\textit{Axial Energy}\\
West Virginia, USA \\
0009-0005-3706-6761}
\and
\IEEEauthorblockN{4\textsuperscript{th} NagaBabu Koganti}
\IEEEauthorblockA{\textit{Power Electronic Controls Engineer} \\
\textit{John Deere}\\
North Carolina, USA \\
0000-0002-8890-6792}
\and
\IEEEauthorblockN{5\textsuperscript{th} Addisalem Kokob W.}
\IEEEauthorblockA{\textit{Lecturer} \\
\textit{Debre Berhan University}\\
Debre Berhan, Ethiopia\\}
\and
\IEEEauthorblockN{6\textsuperscript{th} Praveen Damacharla}
\IEEEauthorblockA{\textit{Research Scientist} \\
\textit{KINETICAI}\\
The Woodlands, Texas, USA \\
0000-0001-8058-7072}
}
\begin{document}
\maketitle
	
\begin{abstract}
This paper presents a multi-domain model for simulating direct current (DC) arcs by coupling thermal, fluid dynamics, and electromagnetic field equations. The magneto-hydrodynamics (MHD) equations are first simplified to derive a two-dimensional (2D) axisymmetric model.  The model is then discretized and implemented in MATLAB/Simulink to capture the spatial and temporal evolution of temperature, velocity, and magnetic fields within the arc column, surrounded by a stabilizing wall. Simulation results for arc currents in the range of 10–1500~A shows that the temperature distribution is strongly radial, with peak values along the central axis and rapid decay toward the surrounding walls. The velocity and magnetic field plots follow the expected physical patterns, validating the accuracy of the approach. Furthermore, the inverse current–voltage characteristic obtained aligns with classical arc models and experimental observations, while transient simulations confirm rapid voltage stabilization within milliseconds. The proposed model thus provides an approximate and simple-to-use model useful for studying DC arc behavior inside DC circuits and offers a foundation for extending arc simulations to system-level applications such as circuit breakers, welding systems, and plasma devices.
\end{abstract}
	
\begin{IEEEkeywords}
DC arcs; magnetohydrodynamics; two-dimensional axisymmetric; Simulink; plasma devices.
\end{IEEEkeywords}
\section{Introduction}
An electric arc is a type of plasma phenomenon that occurs when an electrical current passes through a conducting medium such as air, gas, or plasma. Electric arcs are widely utilized in industrial processes such as welding, circuit breakers, high-intensity lighting, and plasma torches. Accurate modeling of arc discharges is crucial for predicting arc behavior, detecting arc faults, optimizing design, and improving reliability of systems, where arcs are either beneficial (e.g., welding, plasma cutting) or detrimental (e.g., electrical faults, insulation breakdown) \cite{li2013series,lu2020dc}.

Traditionally, DC arc models have been developed using simplified empirical or semi-empirical approaches. These models often describe the arc column as a one-dimensional or quasi-one-dimensional system, employing assumptions of local thermodynamic equilibrium (LTE) and uniform transport properties \cite{taconelli2024overview}. While such models are computationally efficient, they fail to capture the complex interactions between thermal, fluid dynamic, and electromagnetic fields that govern the behavior of arc plasma.

To address these limitations, multi-domain models have been developed, which couple fluid dynamics, heat transfer, and electromagnetics within a self-consistent magneto-hydrodynamics (MHD) framework. In such models, the governing equations consist of the continuity, momentum, and energy conservation laws of the plasma, combined with Maxwell’s equations and Ohm’s law \cite{rau20163, xu20192}. This approach enables the prediction of spatially resolved plasma properties such as temperature, velocity, and magnetic fields, providing insights into arc column dynamics that cannot be obtained from purely empirical models. 

Despite their advantages, multi-domain models present significant challenges. The equations are highly nonlinear, strongly coupled, and often stiff, requiring advanced numerical techniques such as finite element analysis (FEA), and robust time-stepping schemes. Moreover, accurate arc modeling requires reliable thermodynamic and transport property data for the plasma, which must be derived from experimental measurements or detailed kinetic models. Ongoing research continues to refine these models, aiming to strike a balance between physical fidelity with computational efficiency for real-world applications.

This work develops a 2D axisymmetric multi-physics DC arc model formulated in cylindrical coordinates, incorporating temperature-dependent material properties for electrical conductivity $\sigma(T)$, thermal conductivity $k(T)$, and dynamic viscosity $\mu(T)$. The MHD equations are coupled with an explicit induction equation for the azimuthal magnetic field $B_{\phi}$, ensuring self-consistent electromagnetic interaction. The whole system is transformed into a compact state-space representation suitable for MATLAB/Simulink implementation under an algebraic current-closure constraint. This simulation enables system-level studies of arc behavior, including voltage–current characteristics and transient dynamics, while maintaining a physics-consistent description of the internal plasma fields. 

This study focuses on DC arcs occurring in low- and medium-voltage power systems (typically below $1.5$ kV DC) such as photovoltaic arrays, electric vehicle DC buses, and DC microgrids. The analysis specifically addresses wall-stabilized arcs formed between stationary cylindrical electrodes in an air-filled enclosure at $1$ atm. The simulated gap length ranges from $5$ to $100$ mm, with a bore radius of $R=4$ mm, representative of confined arcs in industrial contactors. 
	
\section{Full Magnetohydrodynamics (MHD) Equations}
\begin{figure}
	\begin{center}
		\begin{tikzpicture}[>=Stealth, line width=0.9pt]		
			\draw[dashed, ->] (0,-2) -- (0,-0.5) node [above] {$z$};
			\draw[dashed, ->] (0,-2) -- (0.45,-2) node [right] {$r$};
			
			\draw[thick] (-0.8,-4) rectangle (0.8,-3);
			\node at (0,-3.5) {Anode};
			
			\draw[thick] (-0.8,3) rectangle (0.8,4);
			\node at (0,3.5) {Cathode};
			
			\draw[thick, rounded corners=6pt] (-0.8,-2.9) -- (-0.8,2.9) -- (0.8,2.9) -- (0.8,-2.9) -- cycle;
			
			\draw[->, thick] (0, 1.5) -- (0, 2.5) node[midway, right] {$I_{\text{arc}}$};
			
			\draw[<->] (1.2,-3) -- (1.2,3) node[midway, right] {$V_{arc}$};
			
			\draw[dashed] (-0.8,-2.9) -- (-1.5,-2.9);
			\draw[dashed] (-0.8,-4.0) -- (-1.5,-4.0);
			\draw[<->] (-1.4,-2.9) -- (-1.4,-4.0);
			\node[left] at (-1.6,-3.4) {Anode Region};
			
			\draw[dashed] (-0.8,4.0) -- (-1.5,4.0);
			\draw[dashed] (-0.8,2.9) -- (-1.5,2.9);
			\draw[<->] (-1.4,2.9) -- (-1.4,4.0);
			\node[left] at (-1.6,3.5) {Cathode Region};
			
			\node at (-2.6,0) {Plasma Column};
		\end{tikzpicture}
		\caption{Geometry of studied arc.}
		\label{fig:geometry-of-studied-arc}
	\end{center}
\end{figure}
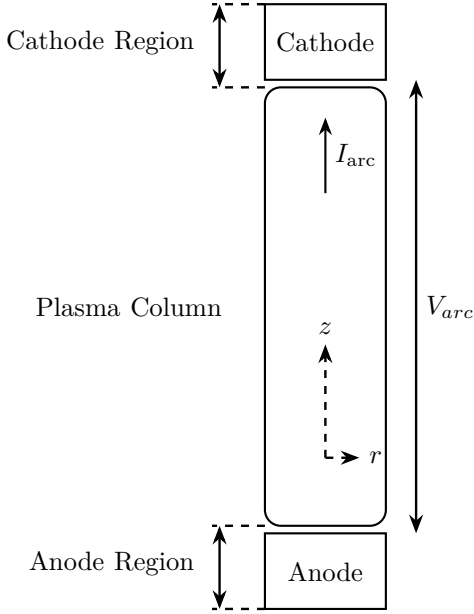

DC electric arcs exhibit complex multi-physics behavior governed by MHD interactions. Traditional 2D/3D MHD models \cite{xu20192,trelles2017finite} are computationally intensive for system-level studies. This work develops a simplified 2D axisymmetric model, as shown in Fig. \ref{fig:geometry-of-studied-arc}, that strikes a balance between physical fidelity with implementation practicality in Simulink. 
The governing MHD equations for a conducting fluid (plasma) are provided below \cite{londero2021analysis}. To describe the behavior of an arc plasma, these equations have to be solved within the arc's occupied region of space. The fully coupled MHD equations for an electrically conducting fluid (plasma) are:
\begin{itemize}
	\item {Mass conservation:}
	\begin{equation}\label{eq:mass-conservation-equation}
		\frac{\partial \rho_v}{\partial t} + \nabla \cdot (\rho_v \mathbf{u}) = 0
	\end{equation}
		
	\item{Energy conservation:}
	\begin{equation}\label{eq:energy-conservation-equation}
		\rho_v c_p \left( \frac{\partial T}{\partial t} + \mathbf{u} \cdot \nabla T \right) = \nabla \cdot (k \nabla T) + \mathbf{J} \cdot \mathbf{E} - q_{\mathrm{rad}}
	\end{equation}
		
	\item {Momentum conservation:}
	\begin{equation}\label{eq:momentum-conservation-equation}
		\rho_v \left( \frac{\partial \mathbf{u}}{\partial t} + (\mathbf{u} \cdot \nabla) \mathbf{u} \right) = - \nabla p + \mathbf{J} \times \mathbf{B} + \nabla \cdot \boldsymbol{\tau}
	\end{equation}
		
	\item {Maxwell's equations (quasi-static approximation):}
	\begin{equation}
		\nabla \times \mathbf{B} = \mu_0 \mathbf{J}, \quad \nabla \times \mathbf{E} = - \frac{\partial \mathbf{B}}{\partial t}
	\end{equation}
		
	\item {Ohm's law:}
	\begin{equation}
		\mathbf{J} = \sigma ( \mathbf{E} + \mathbf{u} \times \mathbf{B} )
	\end{equation}
\end{itemize}
The list of symbols used in the above equations is provided in Table \ref{tab:list of parameters and symbols}.
\begin{table*}
	\centering
	\renewcommand{\arraystretch}{1.2}
	\caption{List of Symbols}
	\label{tab:list of parameters and symbols}
	\begin{tabular}{|c|p{6cm}|c|p{6cm}|}
		\hline
		\textbf{Symbol} & \textbf{Name and Unit} & \textbf{Symbol} & \textbf{Name and Unit} \\
		\hline
		$\rho_v$ & Mass density of the fluid (kg/m$^3$) & $\mathbf{u}$ & Fluid velocity vector (m/s) \\
		$p$ & Pressure (Pa) & $T$ & Temperature (K) \\
		$c_p$ & Specific heat capacity at constant pressure (J/(kg·K)) & $q_{\mathrm{rad}}$ & Volumetric radiative heat loss (W/m$^3$) \\
		\hline
		$\mathbf{E}$ & Electric field vector (V/m) & $\mathbf{B}$ & Magnetic flux density (T) \\
		$\mathbf{J}$ & Current density vector (A/m$^2$) & $\mu_0$ & Magnetic permeability of free space ($4\pi \times 10^{-7}$ H/m) \\
		\hline
		$\sigma$ & Electrical conductivity (S/m) & $k$ & Thermal conductivity (W/(m·K)) \\
		$\mu$ & Dynamic viscosity (Pa·s) & $\boldsymbol{\tau}$ & Viscous stress tensor \newline $\tau_{ij} = \mu \left( \frac{\partial u_i}{\partial x_j} + \frac{\partial u_j}{\partial x_i} - \tfrac{2}{3}\delta_{ij} \nabla \cdot \mathbf{u} \right)$ \\
		\hline
	\end{tabular}
\end{table*}

\section{Solution Approach}
Solving the above set of equations typically requires the use of FEA. However, reasonable simplifying assumptions can reduce computational complexity while preserving the underlying physics. These assumptions mainly depend on the specific physical aspects being modeled. A common assumption in the literature is to treat the arc as an axisymmetric phenomenon, allowing azimuthal variations to be neglected without significant loss of accuracy. In this study, the arc is wall-stabilized and cylindrical, with its axis aligned along the $z-$axis of the cylindrical coordinate system. The use of cylindrical coordinates is self-evident from the assumed shape of an arc.

\subsection{Modeling Assumptions}
In the axisymmetric MHD, the following assumptions were used:
\begin{enumerate}
	\item Axisymmetric ($\partial/\partial\phi=0$) for cylindrical coordinates $(r,\phi,z)$.
	\item Solve for the axial velocity $u_z(r,z,t)$ as the only velocity state; radial velocity $u_{r}$ is computed algebraically from continuity (so it is not an independent integrator state).
	\item Low-Mach and incompressible approximation: $\rho_v$ constant.
	\item Magnetic field: keep the azimuthal component $B_{\phi}(r,z,t)$ as the dynamic magnetic variable.
	\item Ohm's law: neglect motional term $\mathbf u\times\mathbf B$ in Ohm's law.
	\item Axisymmetric Ampere relation used to compute $J_z$ from $B_\phi$:
	\[
	J_z(r,z,t)=\frac{1}{\mu_0 r}\frac{\partial}{\partial r}\big(r B_\phi(r,z,t)\big).
	\]
	\item Lorentz axial force is approximated from radial currents:
	\[
	J_r(r,z) \approx -\frac{r}{2}\frac{\partial J_z}{\partial z},\quad f_{L,z}=J_r B_\phi.
	\]
	\item Thermal physics: $\sigma=\sigma(T)$, $k=k(T)$, $\eta=1/\sigma$. Joule heating is $Q_J = \eta J_z^2 = J_z^2/\sigma$.
	\item Pressure: we treat axial pressure gradient $-\partial p/\partial z$ as either (a) specified input or (b) approximated by a simple model.
\end{enumerate}
	
\section{Simplification of MHD Equations}
With the above assumptions, the main partial differential equations of MHD can be simplified for our case as follows:
\subsection*{Mass Conservation}
\begin{equation}\label{eq:simplified-mass-conservation-equation}
	\nabla \cdot \mathbf{u} = 0
\end{equation}
\subsection*{Energy Conservation}
\begin{align}
	\rho_v c_p&\left(\pderiv{T}{t} + u_r\pderiv{T}{r} + u_z\pderiv{T}{z}\right)
	= \frac{1}{r}\pderiv{}{r}\!\Big(r\,k(T)\pderiv{T}{r}\Big) \nonumber  \\
	&+ \pderiv{}{z}\!\Big(k(T)\pderiv{T}{z}\Big) + \frac{J_z^2}{\sigma(T)} \label{eq:simplified-energy-conservation-equation}
\end{align} 
\subsection*{Momentum Conservation}
\begin{align}
	\rho_v &\left(\pderiv{u_z}{t} + u_r\pderiv{u_z}{r} + u_z\pderiv{u_z}{z}\right)
	= -\pderiv{p}{z} + f_{L,z}(r,z,t) \nonumber \\
	&+ \frac{1}{r}\pderiv{}{r}\!\Big(r \mu(T)\pderiv{u_z}{r}\Big) + \pderiv{}{z}\!\Big(2 \mu(T)\pderiv{u_z}{z}\Big) \label{eq:final-cont-momentum-equation}
\end{align}
\subsection*{Induction (Azimuthal Component $B_\phi$)}
\begin{align}
		\mu_0\frac{\partial B_\phi}{\partial t}
		&= \frac{1}{r} \frac{\partial}{\partial r} \left[r \, \eta(T) \frac{\partial B_\phi}{\partial r} \right] \nonumber \\
		&+ \frac{\partial}{\partial z} \left[ \eta(T) \frac{\partial B_\phi}{\partial z} \right] - \frac{\eta(T)}{r^2} B_\phi \label{eq:simplied-induction-equation}
\end{align}
\subsection*{Auxiliary Algebraic Relations}\begin{equation}
	u_r(r,z,t) = -\frac{1}{r}\int_0^r s\,\pderiv{u_z(s,z,t)}{z}\,ds
\end{equation}
At the wall, the conductive heat flux from the plasma is balanced by the net radiative emission according to the Stefan–Boltzmann law:
\[
-\,k(T)\,\frac{\partial T}{\partial r}\Big|_{r=R}
= \varepsilon\,\sigma_{\mathrm{SB}}\!\left[T^4(R,z,t) - T_w^4\right],
\]
where $\varepsilon$ is the wall emissivity (typically between 0.7 and 0.9), $\sigma_{\mathrm{SB}} = 5.67\times10^{-8}\,\mathrm{W\,m^{-2}\,K^{-4}}$ is the Stefan–Boltzmann constant, and $T_w$ is the wall temperature maintained at $300~\mathrm{K}$.
This boundary formulation replaces the volumetric radiative term in the energy conservation equation and ensures that energy loss occurs only at the plasma–wall interface, consistent with a wall-stabilized arc configuration.
\section{Model Conversion to State-Space Form}
From the 2D axisymmetric $(r,z)$ Partial Differential Equations (PDEs), we can build a time-dependent state-space form that can be implemented in Simulink. We first discretize in $r$ and $z$ (finite-difference/finite-volume style) and give the explicit nodewise Ordinary Differential Equations (ODEs) for each state: temperature $T_{i,j}(t)$, axial velocity $u_{i,j}(t)$, and azimuthal magnetic field $B_{\phi,i,j}(t)$, as in Fig. \ref{fig:discretization-approach}. The following modeling choices are used to facilitate the implementation of simplified equations into state-space form.
\subsection{Key Modeling Choices}
\begin{enumerate}
	\item{Input}: $I(t)$ is the total axial current through the arc (A), given to the model each time.
	\item{Electric field per axial cell} (algebraic): in each axial cell $j$, we compute $E_{z,j}(t)$ from the constraint that the total current crossing that cell equals $I(t)$. This enforces series-arc current.
	\item Joule heating and magnetic sources are fully computed from local fields that depend on $T$ and $I$.
	\item We retain the time-dependent induction PDE for $B_\phi$ (so magnetic diffusion transients are captured). That PDE is discretized and becomes ordinary differential equations (ODEs) for each $B_{i,j}$.
\end{enumerate}
The plasma is assumed to be optically thin, in local thermodynamic equilibrium (LTE), and fully characterized by temperature-dependent transport and electrical properties. The model neglects electrode sheath dynamics, as well as the effects of ablation and erosion. The electrical source is considered stiff, maintaining a prescribed arc current, \(I(t)\). Under these assumptions, the model is valid for investigating steady and transient plasma column behavior and its macroscopic voltage response.
\subsection{Discretization Grid and Notation}
\begin{figure}
	\begin{center}
		\begin{tikzpicture}[scale=1.0]
		
		\def\R{3}   
		\def\H{8}   
		\def\sep{0.1} 
		
		\draw[thick] (\R/2,0) -- (0,0) -- (0,\H) -- (\R/2,\H);
		\draw[thick, rounded corners=6pt] (\R/2,0) -- (\R,0) -- (\R,\H) -- (\R/2,\H);
		
		\draw[thick] (0.0,-0.75) -- (0.0,-\sep) -- (\R,-\sep) -- (\R,-0.75);
		\node at (\R/2,-0.5) {Anode};
		
		\draw[thick] (0.0,\H+0.75) -- (0.0,\H+\sep) -- (\R,\H+\sep) -- (\R,\H+0.75);
		\node at (\R/2,\H+0.5) {Cathode};
		
		\draw[->,dashed] (0,\H+0.25) -- (0,\H+1.25) node[above]{$z$};
		\draw[->,dashed] (0,0) -- (\R+1.0,0) node[right]{$r$};
		
		\def\Nr{6}   
		\def\Nz{10}  
		
		\foreach \i in {1,...,\numexpr\Nr-1} {
			\pgfmathsetmacro{\x}{\i*\R/\Nr}
			\draw[dashed,gray] (\x,0) -- (\x,\H);
		}
		
		\foreach \j in {1,...,\numexpr\Nz-1} {
			\pgfmathsetmacro{\y}{\j*\H/\Nz}
			\draw[dashed,gray] (0,\y) -- (\R,\y);
		}
		
		\fill[green!10,opacity=0.4] (0,0) rectangle (\R,\H);
		
		\draw[red,thick] (0.5,2.4) rectangle (1,3.2);
		\node[red] at (0.75,2.8) {\tiny $(i,j)$};
		
	\end{tikzpicture}
	\caption{Discretization of the arc column.}
	\label{fig:discretization-approach}
	\end{center}
\end{figure}
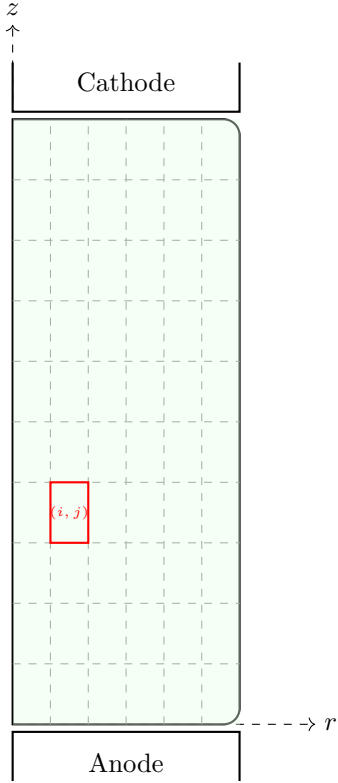

As shown in Fig. \ref{fig:discretization-approach}, we use a cell-centered grid ($N_r$ as radial nodes, $N_z$ as axial cells). Notation: node $(i,j)$ indicates radial index $i=1..N_r$, axial index $j=1..N_z$. Radial spacing is denoted by $\Delta r$, axial spacing by $\Delta z$, and radial node radii by $r_i$. The state vectors are stacked column-wise (all $r$ for $z=1$, $z=2$, and so on). 
For this grid, we use the following differentiation discretization along with their notations:
\begin{itemize}
	\item Radial nodes $r_i$, $i=1,\dots,N_r$ with spacing $\Delta r$
	\item Axial nodes $z_j$, $j=1,\dots,N_z$, spacing $\Delta z$
\end{itemize}
Finite-difference shorthand:
\begin{itemize}
	\item Radial first difference: $\delta_r^+ f_{i,j} = (f_{i+1,j}-f_{i,j})/\Delta r$
	\item Radial centered second: $\delta_{rr} f_{i,j} \approx (f_{i+1,j}-2f_{i,j}+f_{i-1,j})/\Delta r^2$
\end{itemize}
Thus, at each node, we calculate:
\begin{itemize}
	\item $T_{i,j}(t)$ — temperature at each node ($N_r\times N_z$ states),
	\item $u_{i,j}(t)$ — axial velocity $u_z$ at each node ($N_r\times N_z$ states),
	\item $B_{i,j}(t)$ — azimuthal field $B_\theta$ at each node ($N_r\times N_z$ states).
\end{itemize}
with the total number of states becoming $N_x = 3 N_r N_z$. Additionally, for each node, algebraic relations provide  $\sigma_{i,j}=\sigma(T_{i,j})$, $\eta_{i,j}=1/\sigma_{i,j}$, $k_{i,j}=k(T_{i,j})$, $\mu_{i,j}=\mu(T_{i,j})$. 
\subsection{Discrete ODEs}
Discretizing the ODEs using the above approach, we obtain the following:
\subsubsection{Temperature ODE (for interior node $i,j$)} (\ref{eq:simplified-energy-conservation-equation}) when discretized becomes:
\begin{align}
	&\rho_v c_p \,\deriv{T_{i,j}}{t} =  \dfrac{J_{z,i,j}^2}{\sigma_{i,j}} - \rho_v c_p\Big(u_{r,i,j}\,\partial_r^{\rm fwd}T_{i,j} + u_{z,i,j}\,\partial_z^{\rm fwd}T_{i,j}\Big) \nonumber \\
	&+\underbrace{\frac{1}{r_i \Delta r}\Big(r_{i+\tfrac12}\,k_{i+\tfrac12,j} \, \partial_r^{\rm fwd}T_{i,j}
	- r_{i-\tfrac12}\,k_{i-\tfrac12,j}\,\partial_r^{\rm bwd}T_{i,j} \Big)}_{\text{radial conduction}} \nonumber \\
	&+\underbrace{\frac{1}{\Delta z} \Big(k_{i,j+\tfrac12} \,\partial_z^{\rm fwd}T_{i,j} - k_{i,j-\tfrac12}\,\partial_z^{\rm bwd}T_{i,j}\Big)}_{\text{axial conduction}} - q_{\rm rad}(T_{i,j}) \label{eq:discret-energy-equation}
\end{align}
where $\,\partial_r^{\rm fwd}T_{i,j}= \frac{T_{i+1,j}-T_{i,j}}{\Delta r}$, $\,\partial_r^{\rm bwd}T_{i,j}= \frac{T_{i,j}-T_{i-1,j}}{\Delta r}$, $\,\partial_z^{\rm fwd}T_{i,j}= \frac{T_{i,j+1}-T_{i,j}}{\Delta z}$, $\,\partial_z^{\rm bwd}T_{i,j}= \frac{T_{i,j}-T_{i,j-1}}{\Delta z}$. 
All conduction interface coefficients are arithmetic averages, e.g.  $r_{i+\tfrac12} = \left(r_{i+1}+r_i\right)/2$, $r_{i-\tfrac12} = \left(r_{i+1}-r_i\right)/2$, $k_{i,j+\tfrac12}=(k_{i,j+1}+k_{i,j})/2$ and $k_{i,j-\tfrac12}=(k_{i,j}+k_{i,j-1})/2$. We use upwind differencing for the convective derivatives to ensure stability in solutions.
\subsubsection{Axial momentum ODE (for $u_{i,j}$)}
The discretization of (\ref{eq:final-cont-momentum-equation}) gives us:
\begin{align}
	&\rho_v\,\deriv{u_{i,j}}{t} = f_{L,i,j}- \rho_v\Big(u_{r,i,j}\,\partial_r^{\rm fwd}u_{i,j} + u_{z,i,j}\,\partial_z^{\rm fwd}u_{i,j}\Big) \nonumber \\
	&+\underbrace{\frac{1}{r_i\Delta r}\Big(r_{i+\tfrac12}\,\mu_{i+\tfrac12,j}\partial_r^{\rm fwd}u_{i,j} - r_{i-\tfrac12}\,\mu_{i-\tfrac12,j}\partial_r^{\rm bwd}u_{i,j}\Big)}_{\text{radial viscous}} \nonumber \\
	&+ \underbrace{\frac{1}{\Delta z} \Big(\mu_{i,j+\tfrac12}\partial_z^{\rm fwd}u_{i,j} - \mu_{i,j-\tfrac12}\partial_z^{\rm bwd}u_{i,j} \Big)}_{\text{axial viscous}}- G_{i,j}(t) \label{eq:discret-momentum-equation}
\end{align}
where $\partial_r^{\rm fwd}u_{i,j}=\frac{u_{i+1,j}-u_{i,j}}{\Delta r}$, $\partial_z^{\rm fwd}u_{i,j} = \frac{u_{i,j+1}-u_{i,j}}{\Delta z}$, $\partial_r^{\rm bwd}u_{i,j}=\frac{u_{i,j}-u_{i-1,j}}{\Delta r}$, $\partial_z^{\rm bwd}u_{i,j} = \frac{u_{i,j}-u_{i,j-1}}{\Delta z}$. Similar to the energy equation, arithmetic averages are also used to the dynamic viscosity $\mu$.

The axial Lorentz body force density (per volume) uses the approximation from earlier derivations, where 
\begin{equation}
	f_{L,i,j} = J_{r,i,j}\, B_{i,j}
\end{equation}
using 
\begin{equation}
	J_{r,i,j} \approx -\frac{r_i}{2}\,\frac{J_{z,i,j+1}-J_{z,i,j-1}}{2\Delta z}
\end{equation}
for approximate radial current density. $ G_{i,j}(t)$ in the above equation represents the external pressure gradient, which can be taken as a model input.
\subsubsection{Induction ODE (for $B_{i,j}=B_\phi$)}
Discretizing the resistive induction equation (\ref{eq:simplied-induction-equation}) becomes an ODE for each $B_{i,j}$ as:
\begin{align}
	&\mu_0\deriv{B_{i,j}}{t} = \frac{1}{\Delta z}\Big(\eta_{i,j+\tfrac12}\partial_z^{\rm fwd}B_{i,j}
	- \eta_{i,j-\tfrac12}\partial_z^{\rm bwd}B_{i,j} \Big) \nonumber \\
	&+\frac{1}{r_i\Delta r}\Big( r_{i+\tfrac12}\,\eta_{i+\tfrac12,j}\partial_r^{\rm fwd}B_{i,j}
	- r_{i-\tfrac12}\,\eta_{i-\tfrac12,j}\partial_r^{\rm bwd}B_{i,j} \Big) \nonumber \\
	&- \frac{\eta_{i,j}}{r_i^2} B_{i,j} \label{eq:discret-induction-equation}
\end{align}
where $\partial_r^{\rm fwd}B_{i,j}=\frac{B_{i+1,j}-B_{i,j}}{\Delta r}$, $\partial_z^{\rm fwd}B_{i,j} = \frac{B_{i,j+1}-B_{i,j}}{\Delta z}$, $\partial_r^{\rm bwd}B_{i,j}=\frac{B_{i,j}-B_{i-1,j}}{\Delta r}$, $\partial_z^{\rm bwd}B_{i,j} = \frac{B_{i,j}-b_{i,j-1}}{\Delta z}$, $\eta_{i,j}=1/\sigma_{i,j}$ and interface $\eta_{i+1/2,j}$ are averages computed in similar ways as for $k$ and $\mu$.

\subsection{Discretized Algebraic Relations}
Consider a 2D axisymmetric column in cylindrical coordinates $(r,z)$, partitioned into axial cells indexed by $j$. Within a given axial cell $j$, we adopt  an \emph{axially uniform} electric field across the radius inside that cell,
\begin{equation}
	E_z(r,z_j,t) \approx E_{z,j}(t) \qquad \text{for } r\in[0,R]
\end{equation}
The total current through axial slice $j$ is prescribed to be $I(t)$, and by current continuity in a series arc, it is the same for all $z$. Using the assumption $E_z(r)\equiv E_{z,j}(t)$ inside the slice, we obtain
\begin{equation}
	I(t) \;=\; E_{z,j}(t)\,\underbrace{2\pi\int_0^R \sigma(r,z_j,t)\, r\,dr}_{\displaystyle \mathrm{denom}_j(t)}
\end{equation}
On a radial cell–centered grid $\{r_i\}_{i=1}^{N_r}$ with spacing $\Delta r$ and nodal conductivities $\sigma_{i,j}=\sigma(T_{i,j})$,
the integral is approximated by a conservative Riemann sum:
\begin{align*}
	\mathrm{denom}_j(t) &= 2\pi \sum_{m=1}^{N_r} \sigma_{m,j}(t)\, r_m\, \Delta r \\
	E_{z,j}(t) &= \frac{I(t)}{\mathrm{denom}_j(t)}.
\end{align*}
$\mathrm{denom}_j$ is a \emph{conductivity–weighted cross–section measure}. 
High $\sigma$ (hot, conducting core) increases $\mathrm{denom}_j$ and thus reduces the $E$ needed to carry the prescribed $I$; conversely, when the column is cold (small $\sigma$), $\mathrm{denom}_j$ shrinks and the same $I$ requires a larger $E_{z,j}$. The cell voltage-drop is simply $V_j=E_{z,j}\,\Delta z$, and the arc column voltage is $V_{\mathrm{col}}=\sum_j V_j$. Given $E_{z,j}(t)$, the axial current density at node $(i,j)$ is
\begin{equation}
	J_{z,i,j}(t)=\sigma_{i,j}(t)\,E_{z,j}(t)
\end{equation}
\subsection{Overall Model Algorithm}
The following steps were followed in computing the state derivatives within Simulink. This algorithm was implemented using MATLAB function in Simulink. First, input was defined as: $u_{\mathrm{in}}(t) = I(t)$; and optionally $G_{:,j}(t)$ (axial pressure-gradient), $B_{\rm ext}(z,t)$. Then, the vector state was written as: 
\begin{equation}\label{eq:arc-state-definition}
	X = \begin{bmatrix} T_{:,1} \cdots T_{:,N_z} u_{:,1} \cdots u_{:,N_z} B_{:,1} \cdots B_{:,N_z} \end{bmatrix}^T
\end{equation}
At each time step:
\begin{enumerate}
	\item Compute $\sigma_{i,j}=\sigma(T_{i,j})$ from known temperature dependent formula for plasma conductivity or lookup table of conductivity of plasma with respect to temperature. It is ensured to limit the conductivity in the range $\sigma\ge\sigma_{\min}>0$.
	\item Compute  \emph{conductivity–weighted cross–section measure}, $\mathrm{denom}_j$, and then $E_{z,j} = I(t)/\mathrm{denom}_j$.
	\item Compute $J_{z,i,j}=\sigma_{i,j} E_{z,j}$.
	\item Compute the arc column voltage $V_{\rm column}=\sum_j E_{z,j}\Delta z$.
	\item Compute $B_{i,j}$ used in Joule and Lorentz as the dynamic state from (Bnode). Add $B_{\rm ext}$ at the wall as boundary condition and/or set an initial field computed from Ampère.
	\item Compute $u_{r,i,j}$ from continuity (algebraic integral):
	\begin{equation*}
		u_{r,i,j} = -\frac{1}{r_i}\sum_{m=1}^{i} \Big(\frac{u_{m,j+1}-u_{m,j-1}}{2\Delta z}\Big)\, r_m\,\Delta r .
	\end{equation*}
	\item Compute state derivatives using (\ref{eq:discret-energy-equation}), (\ref{eq:discret-momentum-equation}), and (\ref{eq:discret-induction-equation}) and integrate them to obtain states.
\end{enumerate}
\subsection{Boundary and Numerical Conditions}
The governing equations are discretized in a two-dimensional axisymmetric $(r,z)$ domain with uniform spacing in both directions. The applied boundary and initial conditions, as well as numerical settings, are summarized in Table~\ref{tab:BCs_numerics}.
\begin{table*}[!h]
	\centering
	\caption{Boundary and numerical conditions used in the MHD arc simulation.}
	\renewcommand{\arraystretch}{1.2}
	\begin{tblr}{|l|l|l|}
		\toprule
		\textbf{Location / Parameter} & \textbf{Condition or Value} & \textbf{Description} \\
		\midrule
		$r = 0$ & $\dfrac{\partial T}{\partial r} = 0,\; u_r = 0,\; \dfrac{\partial u_z}{\partial r} = 0,\; \dfrac{\partial B_\phi}{\partial r} = 0$ & Axisymmetry at the centerline \\
		$r = R$ & $T = T_w = 300~\text{K},\; u_z = 0,\; u_r = 0$ & Wall temperature fixed (no-slip, no penetration) \\
		$z = 0,~L$ & Specified pressure gradient $G(z)$ ($=0$ in our case) & Imposed axial driving pressure or flow constraint \\
		$B_\phi$ boundaries & $B_\phi(r=R,z) = B_\text{ext}$,\; $\dfrac{\partial B_\phi}{\partial r}\big|_{r=0} = 0$ & External field at wall, symmetry on axis \\
		\textbf{Initial conditions} & $T(r,z,0) = 300~\text{K}$,\; $u_z(r,z,0) = 0$,\; $B_\phi(r,z,0) = B_\text{ext}$ & Cold-start, quiescent gas \\
		\textbf{Grid resolution} & $(N_r,N_z) = (10,5)$,\; $\Delta r = R/N_r = 0.2~\text{mm}$,\; $\Delta z = L/N_z$ & Uniform spatial discretization \\
		\textbf{Time step} & Adaptive $\Delta t$, solver-controlled & Determined by ODE solver stability (stiff system) \\
		\textbf{Solver} & MATLAB \texttt{ode15s} (stiff, variable-step) & Tolerances: RelTol = $10^{-3}$, AbsTol = $auto$ \\
		\bottomrule
	\end{tblr}
	\label{tab:BCs_numerics}
\end{table*}
\section{Simulation Results}
\subsection{Simulink Model}
\begin{figure}
	\begin{center}
		\subfloat[Simulink arc model \label{fig:simulink-arc-model}]{
			\includegraphics[width=1.0\linewidth]{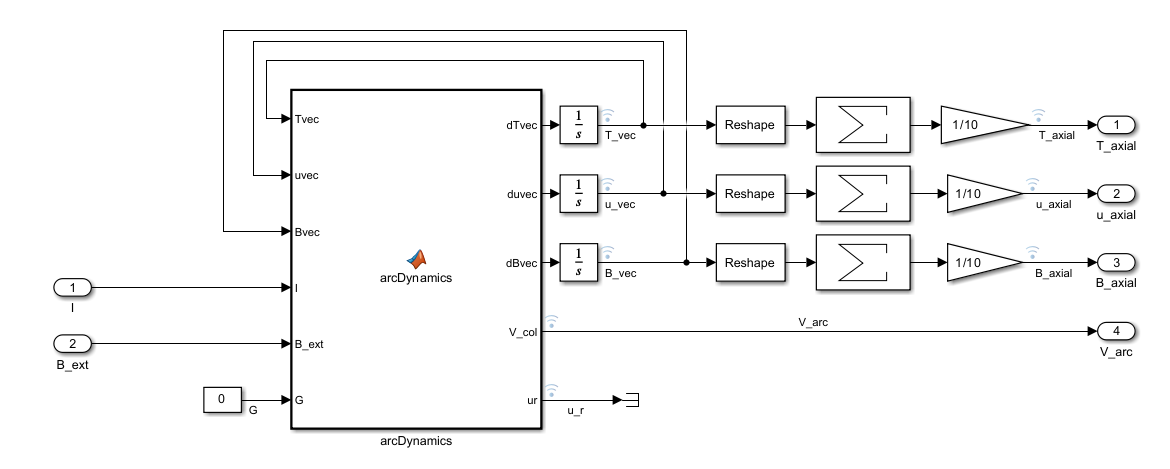}
		}\\
		\subfloat[Arc model used for circuit simulations \label{fig:in-circuit-simulation-arc-model}]{
			\includegraphics[width=1.0\linewidth]{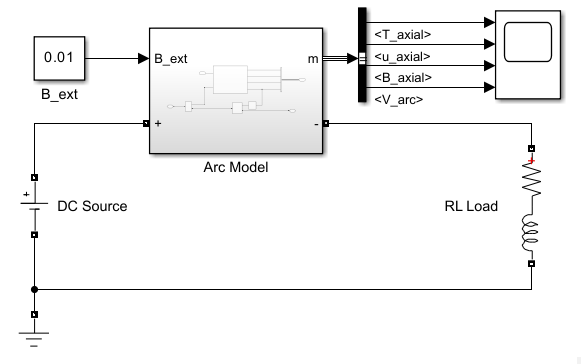}
		}
	\end{center}
	\caption{Simulink arc models.}
	\label{fig:arc-simulink-models-used}
\end{figure}

The implementation of the previously discussed arc model in Simulink is shown in Fig. \ref{fig:simulink-arc-model}. The model takes arc current ($I$) and external magnetic flux density ($B\_ext$) as inputs and outputs arc temperature, axial velocity, axial magnetic flux density, and arc column voltage.  It uses the algorithm, written as a MATLAB function, to compute the time derivative of the arc state ($X$) defined in (\ref{eq:arc-state-definition}). Subsequent Simulink blocks are used after the MATLAB function block to calculate radial averages of arc temperature ($T\_axial$), axial velocity ($u\_axial$), and axial magnetic flux density ($B\_axial$).
Fig. \ref{fig:in-circuit-simulation-arc-model} shows a sample circuit simulation model built using the arc model to study arc properties within an R-L circuit using the Simulink\textit{ Specialized Power Systems (sps)} toolbox components. The arc is connected in series with the load. This model can be used in other power system simulations to simulate the effect of series DC arc.
\subsection{Two-Dimensional Results}
\begin{figure}
	\begin{center}
		\subfloat[Temperature field \label{fig:temperature-field-plot}]{
			\includegraphics[width=0.5\linewidth]{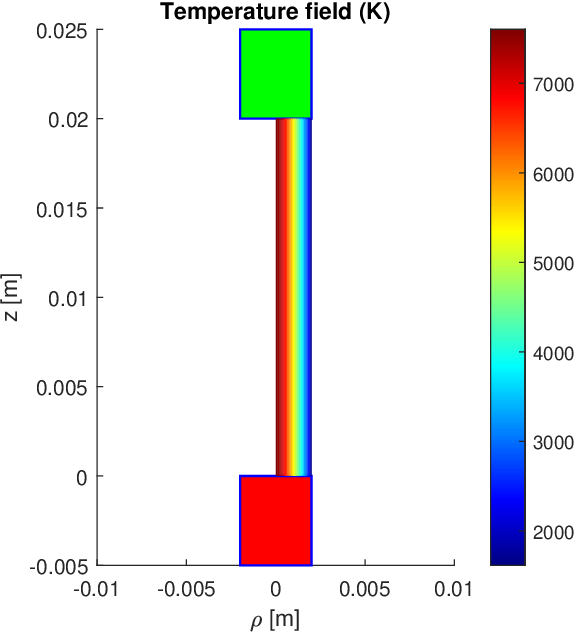}
		}
		\subfloat[Velocity field \label{fig:velocity-field-plot}]{
			\includegraphics[width=0.5\linewidth]{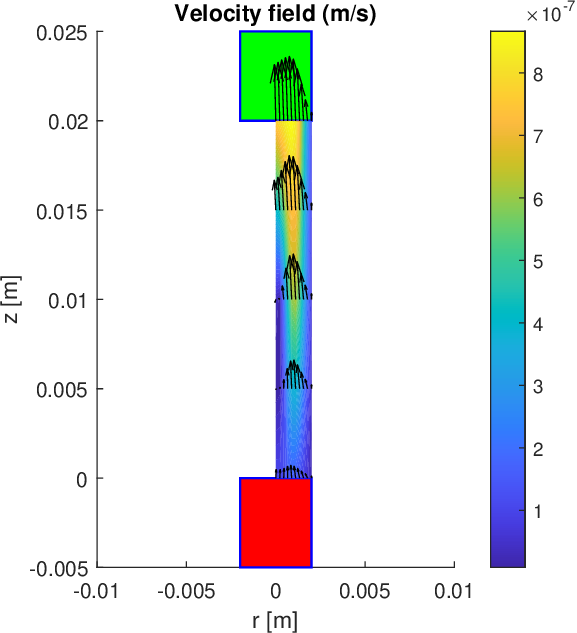}
		}\\
		\subfloat[Magnetic field \label{fig:magnetic-field-plot}]{
			\includegraphics[width=0.5\linewidth]{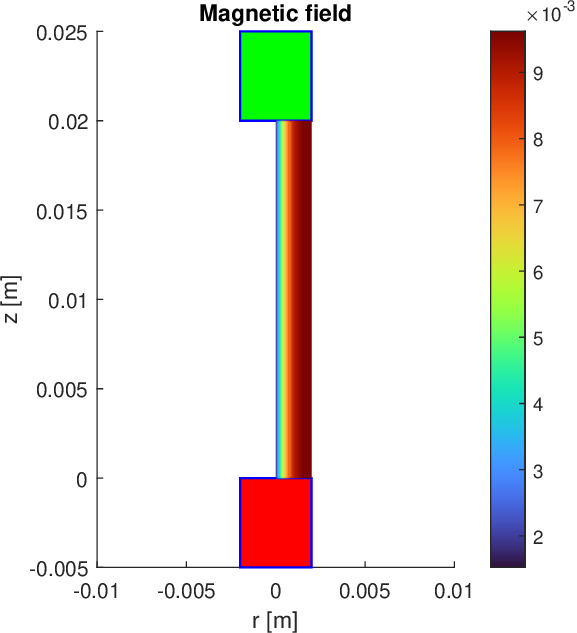}
		}
	\end{center}
	\caption{Two-dimensional temperature and velocity field plots.}
	\label{fig:two-dimensional-temperature-and-velocity-field-plots}
\end{figure}

In reference to the model in Fig. \ref{fig:arc-simulink-models-used}, Fig. \ref{fig:temperature-field-plot} shows temperature,  Fig. \ref{fig:velocity-field-plot} shows velocity, and Fig. \ref{fig:magnetic-field-plot} depicts azimuthal magnetic field surface plots for an arc current of $I=110$ $A$.  From Fig. \ref{fig:temperature-field-plot}, it is clear that the temperature is hotter near the central axis of the arc column and decreases as we move towards the stabilizing wall. The temperature distribution does not show axial variation. Due to the absence of external pressure gradient, the velocity profile of the plasma, as shown in Fig. \ref{fig:velocity-field-plot}, exhibits no significant variation across the plasma column. 
The magnetic field plot shown in Fig. \ref{fig:magnetic-field-plot} is dominated by the externally applied field used in the simulation.
\subsection{Current-Voltage Relationship}
\begin{figure}[!tb]
	\centering
	\includegraphics[width=0.9\linewidth]{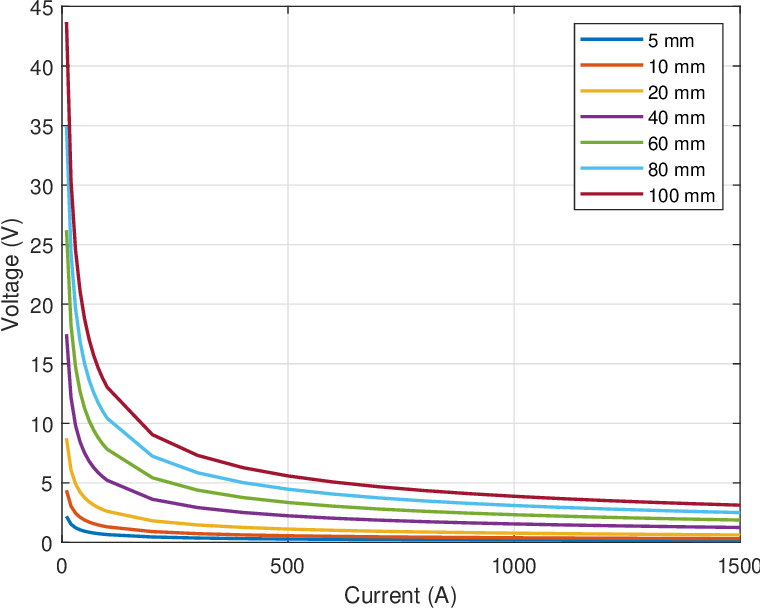}
	\caption{Current versus voltage curve.}
	\label{fig:currentvsvoltagecurve}
\end{figure}

\begin{figure}[!htb]
	\centering
	\includegraphics[width=1.0\linewidth]{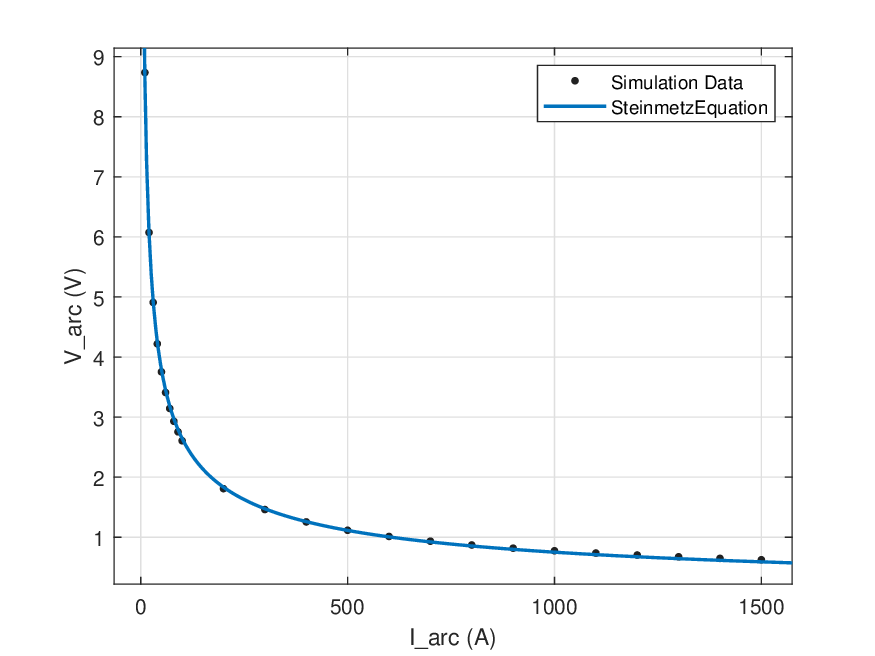}
	\caption{Simulation data fitted to Steinmetz Equation for $L=20\,mm$.}
	\label{fig:steinmetze-quation-curve-fitting}
\end{figure}

\begin{figure}[!tb]
	\begin{center}
		\subfloat[Voltage vs current \label{fig:Voltage-vs-curren}]{
			\includegraphics[width=0.5\linewidth]{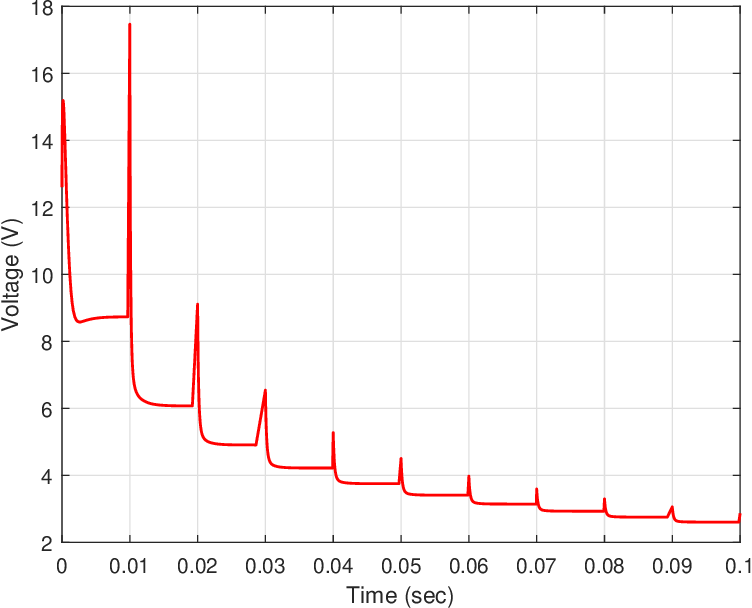}
		}
		\subfloat[Average axial temperature vs current \label{fig:Temperature-vs-current}]{
			\includegraphics[width=0.5\linewidth]{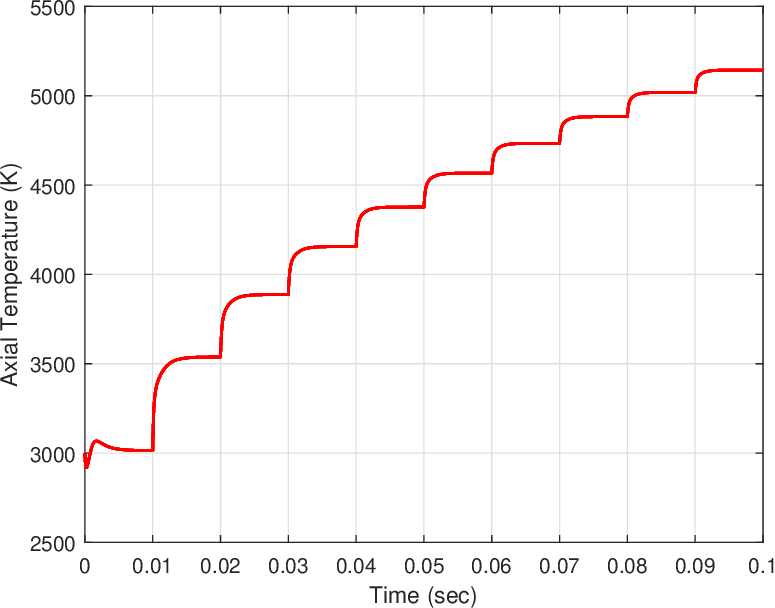}
		}
	\end{center}
	\caption{Variation of arc voltage and average axial temperature with arc current.}
	\label{fig:Voltage-and-temperature-vs-current-plots}
\end{figure}

To study the effect of arc current on the arc column voltage and average temperature distribution, we carried out several simulations of varying arc current in the range $[10, 1500]$ $A$ and different arc lengths in the range $[5, 100] \,mm$. Each current value starting from $I = 10$ $A$ was kept for $10$ $ms$ interval. At the end of this interval, the current was stepped up to the next one. The current versus voltage relationship curve obtained this way is shown in Fig. \ref{fig:currentvsvoltagecurve},
where there is an inverse current-voltage relationship consistent with well-known arc models and experimental data. The effect of increasing arc length is also depicted with the increase in the arc column voltage.
For comparison with known DC arc models, the arc $v-i$ relationship for arc length of $L = 20\,mm$ has been fitted to the Steinmetz Equation model described in \cite{ammerman2009dc}, as shown in Fig. \ref{fig:steinmetze-quation-curve-fitting}. The fitting produced a model with parameters $A = -0.1247$, $C=4.3224$, and $D=5.409$, with an RMSE of $0.0371$.
Fig. \ref{fig:Voltage-vs-curren} shows the arc column voltage as the current is varied from $10$ $A$ to $110$ $A$ at the interval of $10$ $A$. A plot of the average temperature of each axial segment is shown in Fig. \ref{fig:Temperature-vs-current}. One can see that the voltage settles to its final value within $0.01$ s. The final voltage value, after the transient state has passed, shows a similar relationship with the arc current as in Fig. \ref{fig:currentvsvoltagecurve}.
\section{Conclusion}
The simulation results presented in this work demonstrate that the developed multi-domain arc model accurately captures the coupled behavior of thermal, fluid, and electromagnetic fields within the arc column. The predicted temperature, velocity, and magnetic field distributions follow expected physical trends, with peak temperatures along the central axis and reduced gradients toward the wall. Furthermore, the obtained inverse current-voltage characteristic agrees with classical arc models and experimental observations, thereby validating the model's accuracy. Both transient and steady-state responses confirm its reliability for analyzing arc dynamics across different current levels, making it a useful tool for both scientific study and engineering applications involving arc phenomena.
	

	
\end{document}